\documentclass[]{bytedance_seed}

\usepackage[toc,page,header]{appendix}

\usepackage{minitoc}
\usepackage{amsmath,amssymb}
\usepackage{bm}
\usepackage{wrapfig}

\usepackage{enumitem}

\title{ Seed3D 1.0: From Images to High-Fidelity Simulation-Ready 3D Assets}

\affiliation[]{ByteDance Seed}

\abstract{
Developing embodied AI agents requires scalable training environments that balance content diversity with physics accuracy. World simulators provide such environments but face distinct limitations: video-based methods generate diverse content but lack real-time physics feedback for interactive learning, while physics-based engines provide accurate dynamics but face scalability limitations from costly manual asset creation. We present \textbf{Seed3D 1.0}, a foundation model that generates simulation-ready 3D assets from single images, addressing the scalability challenge while maintaining physics rigor. 
Unlike existing 3D generation models, our system produces assets with accurate  geometry, well-aligned  textures, and realistic physically-based materials.
These assets can be directly integrated into physics engines with minimal configuration, enabling  deployment in robotic manipulation and simulation training.
Beyond individual objects, the system scales to complete scene generation through assembling objects into coherent environments.
By enabling scalable simulation-ready content creation, Seed3D 1.0 provides a foundation for advancing physics-based world simulators.
Seed3D 1.0 is now available on \href{https://console.volcengine.com/ark/region:ark+cn-beijing/experience/vision?modelId=doubao-seed3d-1-0-250928&tab=Gen3D}{Volcano Engine}\footnote{Model ID: doubao-seed3d-1-0-250928}.
}

\checkdata[Official Page]{\url{https://seed.bytedance.com/seed3d}}
\correspondence{Jianfeng Zhang (\email{jianfengzhang@bytedance.com})}

\begin{document}
\maketitle
\vspace{-6mm}
\begin{figure}[hb]
    \centering
        \includegraphics[width=\textwidth]{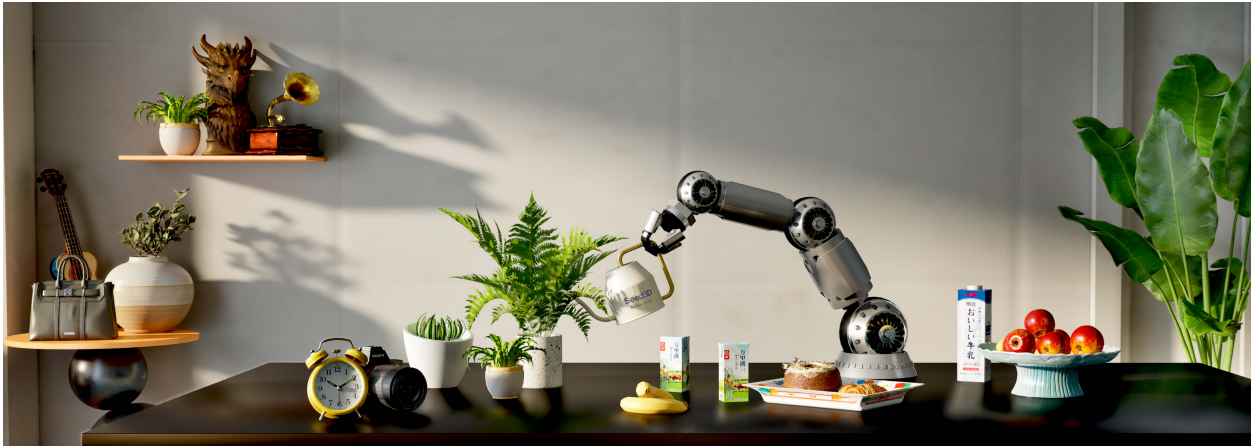}
        \caption{
        \textbf{Seed3D 1.0 generates high-fidelity, simulation-ready 3D assets from single images.} Individual objects generated by our system can be composed into complex scenes for simulation and robotic applications. This kitchen environment demonstrates robotic manipulation simulation with diverse generated assets.
        Best viewed with 8$\times$ zoom.
        }
        \label{fig:teaser}
\end{figure}

\newpage

\tableofcontents
\newpage


\section{Introduction}

Large multimodal models (LMMs) have rapidly evolved from passive chatbots to active agents capable of interacting with tools, APIs, and software environments~\cite{yin2024survey}. This progress advances a longstanding AI objective: building autonomous robots that can perceive, reason, and act in the physical world. However, current LMMs lack essential capabilities for physical interaction—understanding 3D object structure, spatial relationships, material properties, and physical dynamics~\cite{yang2024thinking,zhu2024spa}. A household robot, for instance, must accurately perceive object positions behind partial occlusions, infer material properties for appropriate grasping forces, and predict manipulation consequences in cluttered spaces.

The fundamental challenge is data scarcity. Internet data, while vast, is biased toward text and 2D representations and cannot provide the rich spatial-physical information that embodied systems require. Recent breakthroughs in reinforcement learning, particularly in coding domains where models learn from code execution environments~\cite{le2022coderl,guo2025deepseek} demonstrate how interactive environments can overcome data limitations through structured feedback. However, extending this paradigm to embodied AI demands high-fidelity simulation environments that provide meaningful feedback for spatial reasoning and physical manipulation tasks, which remains largely absent.

Existing world simulators face a fundamental trade-off. Video-based approaches like Cosmos~\cite{agarwal2025cosmos} and Genie-3~\cite{genie3} generate diverse content but lack 3D consistency and the intermediate feedback mechanisms, although these are essential for training embodied agents. Physics-based simulators like IsaacGym~\cite{makoviychuk2021isaac} provide rigorous dynamics and explicit physics modeling for interpretability and safety, but face severe scalability limitations: manual asset creation requires substantial expertise and time, fundamentally constraining the variety and scale of training environments.

To address this challenge, an effective world simulator must bridge content scalability with simulation fidelity. It should generate diverse, physically plausible 3D content while providing faster-than-real-time physics feedback for interactive agent training. In this report, we introduce \textbf{Seed3D 1.0}, a foundation model for simulation-ready 3D asset generation that advances this vision (Figure~\ref{fig:teaser}). Seed3D 1.0 addresses the content scalability challenge by generating high-quality 3D assets that integrate effectively with physics engines, combining generative diversity with simulation rigor. This design preserves explicit physics modeling for interpretability and safety while alleviating the content bottleneck that limits traditional simulation pipelines. Our system demonstrates three key capabilities:
\begin{itemize}
    \item \textbf{High-Fidelity Asset Generation:} 
    Seed3D 1.0 produces 3D assets with detailed geometry,   photorealistic textures (up to 4K resolution), and physically plausible PBR materials that ensure realistic lighting interactions under various illumination conditions. Unlike existing approaches that often produce geometric artifacts or texture misalignment, our model ensures high-quality, consistent assets suitable for both rendering and physical simulation.
    
    \item \textbf{Physics Engine Compatibility:} 
    Assets generated by Seed3D 1.0 integrate seamlessly into physics engines with minimal configuration. We demonstrate practical applications in simulation-based data generation, where these assets create diverse manipulation scenarios for training robotic manipulation models. In addition to data collection, the physics compatibility naturally supports interactive environments for reinforcement learning, where agents acquire skills through environmental feedback.

    \item \textbf{Scalable Scene Composition:} 
    Beyond individual assets, Seed3D 1.0 employs a factorized approach to scene generation: vision-language models understand and plan spatial layouts, while our generative model creates and places assets according to these layouts, enabling coherent scene composition from indoor to urban environments.
    
\end{itemize}

By enabling scalable generation of simulation-ready 3D assets and scene-level composition, Seed3D 1.0 represents a significant step toward practical world simulators. In the following sections, we detail our technical approach and present comprehensive experimental validation demonstrating our system's capabilities across diverse simulation and robotic applications.

\newpage
\section{Model Design}

\subsection{Geometry}
Geometry generation in Seed3D 1.0 focuses on creating high-fidelity, simulation-ready 3D shapes with watertight, manifold geometry, enabling reliable physics simulation while preserving structural details. Similar to 2D generation tasks~\cite{esser2024scaling,rombach2022high}, our approach learns to denoise 3D geometry in a compressed latent space, combining variational autoencoders (VAE)~\cite{kingma2014autoencoding} with rectified flow-based diffusion transformers (DiT)~\cite{liu2023flowstraight}.
This architecture consists of two key components:
\begin{itemize}
    \item \textbf{Seed3D-VAE:} A VAE that learns compact latent representations of 3D geometry, enabling efficient encoding and reconstruction of complex mesh structures while preserving local surface details. 
    \item \textbf{Seed3D-DiT:} A rectified flow-based DiT operating in the learned latent space to synthesize diverse 3D shapes conditioned on reference images.
\end{itemize}

\subsubsection{Seed3D-VAE} 
Seed3D-VAE follows the design of 3DShape2VecSet~\cite{zhang20233dshape2vecset,chen2025dora}, which encodes surface point clouds into a latent vector set and reconstructs continuous geometric representations~\cite{zhang2024clay,hunyuan3d-2.1}. We adopt truncated signed distance functions (TSDFs) as the supervision signal~\cite{chen2025dora}, effectively constraining the regression range while preserving fine details.

\begin{figure}[t]
	\centering
	\includegraphics[width=\columnwidth]{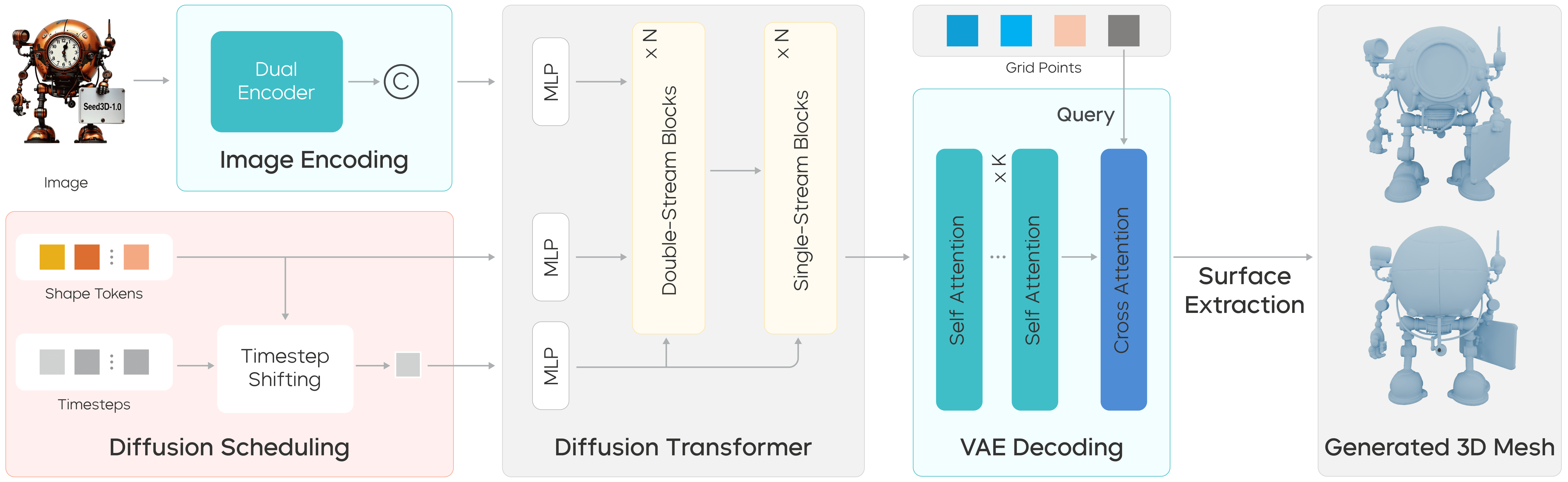}
	\caption[12]{\textbf{Overview of the Seed3D 1.0 geometry generation pipeline.} The framework combines a variational autoencoder named Seed3D-VAE, which is designed for compact geometry encoding and TSDF decoding, with a rectified flow–based Transformer named Seed3D-DiT to generate high-fidelity 3D shapes from input images.} 
	\label{fig:shape_pipeline}
\end{figure}

\textbf{Architecture}. 
Similar to Dora~\cite{chen2025dora}, Seed3D-VAE employs a dual cross-attention encoder and a self-attention decoder~\cite{hunyuan3d-2.0,hunyuan3d-2.1,zhang2024clay,li2024craftsman3d}. Given an input mesh, we uniformly sample points $P_u$ and extract salient edge points $P_s$, which are embedded using Fourier positional encoding~\cite{tancik2020fourier} $\mathrm{PE}(P)$ and concatenated with surface normals $n_P$, where $P = P_u \cup P_s$ denotes the set of combined points. The encoder maps the point set to a compact latent set $\mathbf{Z}=\{\mathbf{z}_m\}_{m=1}^M$ via stacked cross-attention and $L_e$-layer self-attention blocks~\cite{vaswani2017attention}: 
\begin{equation}
    \mathbf{Z}_0 = \mathrm{CrossAttn}(\mathrm{PE}(P), n_P), \quad 
    \mathbf{Z}_i = \mathrm{SelfAttn}(\mathbf{Z}_{i-1}), \; i=1,\dots,L_e.
\end{equation}
The decoder defines a continuous TSDF field $f(x|\mathbf{Z}) : \mathbb{R}^3 \to \mathbb{R}$ conditioned on the latent token set $\mathbf{Z}\in\mathbb{R}^{M\times d}$, mapping query point $x$ to its predicted signed distance value $\hat{d}(x)$.
Specifically, query points $x$ are first embedded with Fourier features $\mathrm{PE}(x)$, refined via $L_k$ self-attention layers, and then attend to latent descriptors $\{\mathbf{z}_m\}$ through cross-attention, followed by an MLP head: 
\begin{equation}
    \hat{d}(x) = \mathrm{MLP}\!\Big(\mathrm{CrossAttn}\big(\mathrm{SelfAttn}^{(j)}(\mathrm{PE}(x)),\, \mathbf{Z}\big)\Big), \; j=1,\dots,L_k.
\end{equation}

\textbf{VAE Training.} 
To enable generalization across different computational budgets and improve robustness, we employ a multi-scale training strategy. We randomly sample token lengths $M \in \{256, 512, \dots, 4096\}$ during training, leveraging the vector set architecture's length-agnostic property—latent tokens are position-encoding-free and permutation-invariant, allowing the decoder to scale beyond token lengths seen during training.
The overall training objective combines TSDF reconstruction loss $\mathcal{L}_{\text{recon}}$ and KL divergence regularization~\cite{kingma2014autoencoding} $\mathcal{L}_{\text{KL}}$:
\begin{equation}
    \mathcal{L}_{\text{VAE}} = \mathcal{L}_{\text{recon}} + \lambda_{\mathrm{KL}} \mathcal{L}_{\text{KL}}
\end{equation}
We employ a warm-up schedule where the KL weight $\lambda_{\mathrm{KL}}$ begins at a small value and gradually increases to its target value ($\lambda_{\mathrm{KL}}=10^{-4}$
) over the course of training to ensure stable convergence.

\subsubsection{Seed3D-DiT} 
Building upon the geometry-aware latent space learned by Seed3D-VAE, Seed3D-DiT employs a rectified flow-based diffusion framework to generate 3D shapes by modeling the transformation from noise to structured latent representations conditioned on image inputs. Below we detail the key architectural components and training procedure.

\textbf{Image Conditioning Module.}  
To capture rich visual semantics for geometry generation, we adopt a dual-encoder design combining DINOv2~\cite{oquab2023dinov2} and RADIO~\cite{radio}. RADIO complements DINOv2 by providing enhanced geometric understanding through knowledge distillation from multiple vision foundation models, helping resolve depth ambiguity in single-view conditioning and improving training stability. Input images are encoded by both networks, and their feature representations are concatenated channel-wise to form comprehensive conditioning signals that capture both semantic and geometric properties.

\textbf{Transformer Architecture.} 
We employ a transformer as the diffusion backbone to model cross-modal relationships between visual and geometric representations. Following the hybrid design of FLUX~\cite{flux2024}, the architecture incorporates double-stream and single-stream processing blocks.
Double-stream blocks process shape and image tokens via modality-specific parameters (distinct layer normalization, QKV projections, and MLPs) while enabling cross-modal interaction via attention on concatenated tokens. Single-stream blocks then process the refined shape tokens through additional transformer layers before final decoding via the Seed3D-VAE decoder. This hybrid approach balances cross-modal learning with modality-specific processing.

\textbf{Diffusion Scheduling.} 
Our training employs a flow matching~\cite{lipman2022flow} framework with velocity field prediction, where timesteps are sampled from a logit-normal distribution. Since longer latent sequences require higher noise levels to effectively disrupt their structure, we apply a length-aware timestep shift~\cite{esser2024scaling} that scales the noise schedule according to sequence length.
During inference, we use deterministic sampling through the learned velocity fields to generate 3D shapes conditioned on input images.

\subsection{Texture}
Beyond 3D shape generation, high-quality texture synthesis is equally critical for creating realistic 3D assets. 
Our texture generation pipeline produces physically-based materials~\cite{burley2012physically} through three sequential components:
\begin{itemize}
    \item \textbf{Seed3D-MV:} A multi-view diffusion model that generates consistent RGB images from multiple viewpoints, conditioned on a reference image and 3D shape guidance.
    \item \textbf{Seed3D-PBR:} A diffusion  model that decomposes multi-view RGB images into albedo, metallic, and roughness maps for physically-based rendering.
    \item \textbf{Seed3D-UV:} A diffusion-based UV inpainting model that addresses self-occlusion artifacts by enhancing texture completeness in UV space.
\end{itemize}

\subsubsection{Seed3D-MV}
Existing multi-view generation works~\cite{Tang2023mvdiffusion,wang2023imagedream} incorporate multi-view attention mechanisms into diffusion models. Though achieving multi-view consistency in image synthesis, these works typically require additional modules such as ControlNet~\cite{zhang2023adding} or MVAdapter~\cite{huang2024mvadapter} to encode geometry and reference image guidance, introducing significant parameter overhead.
Recent work~\cite{liang2025UnitTEX} alleviates this by concatenating multi-view images and computing cross-view attention through fine-tuning pretrained DiT models~\cite{flux2024}. However, this approach can produce suboptimal results when applied to in-the-wild images, as the underlying DiT architecture was not originally designed for multi-view generation.

\begin{figure}[hbtp]
	\centering
	\includegraphics[width=\columnwidth]{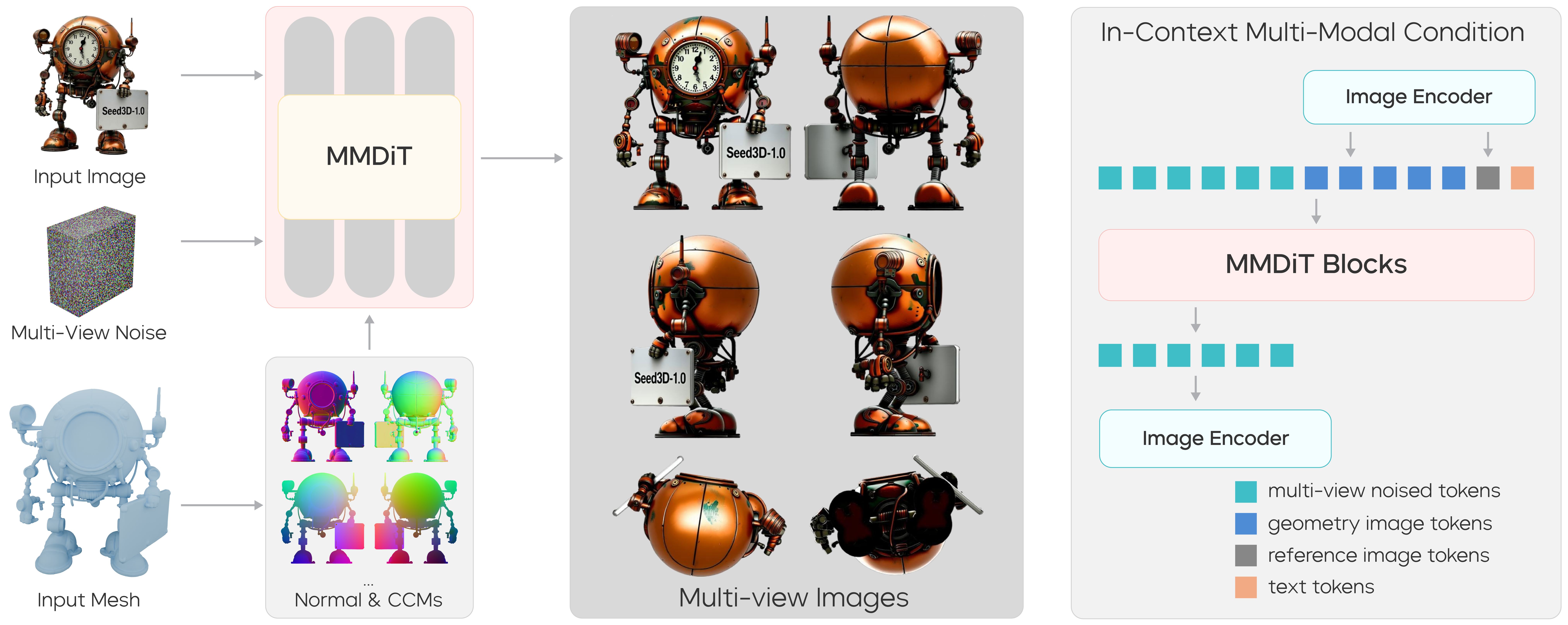}
	\caption{\textbf{Seed3D-MV architecture.} Left: System overview showing the multi-modal conditioning pipeline. Right: In-context multi-modal conditioning mechanism integrating geometry, reference image, and text information.}
	\label{fig:mv_pipeline}
\end{figure}

To address these limitations, we develop Seed3D-MV based on the Multi-Modal Diffusion Transformer (MMDiT) architecture~\cite{esser2024scaling}. As illustrated in Figure~\ref{fig:mv_pipeline}, our approach introduces an in-context multi-modal conditioning strategy with specialized positional encoding. To handle the increased sequence length in multi-view generation, we employ shifted timestep sampling to maintain generation quality.

Our objective is to learn the conditional distribution for multi-view consistent image generation:
\begin{equation}
p(x|g,i,c),
\end{equation}
where $x$ represents the target multi-view images, $g$ denotes spatially aligned multi-view geometry images (\textit{i.e.}, normal maps and canonical coordinate maps rendered from the input mesh), $i$ is the reference image, and $c$ is an optional text prompt.

\textbf{In-Context Multi-Modal Conditioning.} Following UniTex~\cite{liang2025UnitTEX} and Flux.1 Kontext~\cite{labs2025flux1kontextflowmatching}, we enable multi-modal conditioning by concatenating noisy input tokens with clean condition tokens from geometry, reference image, and text modalities along the sequence dimension. This design provides flexible integration of diverse control signals. Specifically, geometry and reference images are encoded into latent representations using a frozen VAE, while text prompts are processed through a pretrained language model~\cite{qwen}. During training, we randomly drop conditional tokens to enable classifier-free guidance~\cite{ho2022classifierfree}.

\textbf{Positional Encoding.} We employ cross-modal RoPE to facilitate interaction between multi-modal tokens. To accommodate the newly introduced token types, we modify the standard RoPE scheme~\cite{su2023rope} to handle both spatially aligned geometry tokens and non-aligned reference image tokens through targeted positional encoding adjustments. Our token sequence is organized as follows: multi-view noisy tokens, geometry image tokens, reference image tokens, and text tokens. This configuration optimizes cross-modal attention while maintaining RoPE compatibility. Empirically, we find that using separate spatial positions for noisy tokens and geometry tokens outperforms shared spatial positioning.

\textbf{Timestep Sampling.} Multi-view generation significantly increases input sequence length, challenging the model's learning capacity and potentially degrading output quality. To maintain high-fidelity generation, we adopt resolution-aware timestep sampling~\cite{esser2024scaling} with a shift-SNR sampling distribution that adapts dynamically based on the noisy token sequence length during both training and inference.

\subsubsection{Seed3D-PBR}
High-quality material generation is essential for realistic 3D content creation. Physically-based rendering (PBR) materials, comprising albedo, metallic, and roughness components, are fundamental for achieving photorealistic rendering results. Existing PBR synthesis methods fall into two categories: generation-based approaches~\cite{he2025materialmvp,kocsis2025intrinsix} that synthesize PBR maps from reference images and 3D geometry, and estimation-based methods~\cite{li2025idarb} that decompose multi-view images directly into material components. Due to limited high-quality PBR training data, generation methods often produce less realistic results compared to estimation approaches. We therefore adopt the estimation paradigm and introduce Seed3D-PBR, which decomposes multi-view images generated by Seed3D-MV into multi-view consistent albedo, metallic, and roughness maps. Unlike existing methods~\cite{he2025materialmvp, li2025idarb, chen2024intrinsicanything, he2025neural}, we propose a DiT-based architecture with a parameter-efficient two-stream design to improve estimation accuracy while handling the distinct characteristics of different material properties.

\textbf{Model Architecture.} Our PBR estimation model is built upon the MMDiT architecture with an innovative two-stream design that enhances alignment between different material modalities (albedo vs. metallic-roughness) while ensuring 3D consistency across viewpoints. The model takes camera pose embeddings, multi-view images, and a reference image as input, and simultaneously generates multi-view albedo and metallic-roughness (MR) maps with cross-view consistency. \par
\begin{figure}[hbtp]
	\centering
	\includegraphics[width=\columnwidth]{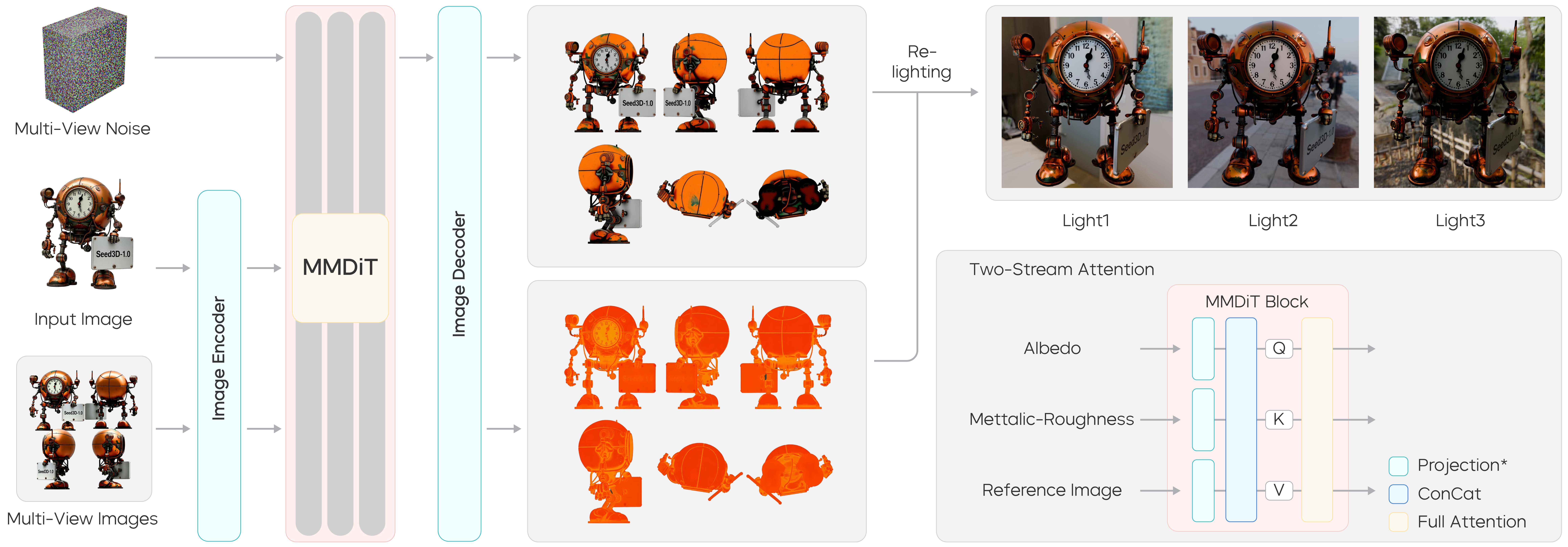}
	\caption{\textbf{The overview of Seed3D-PBR model.} 
    To handle albedo and metallic-roughness in a single DiT model, we propose a network with two-stream attention blocks. \(^*\)Projection contains MLPs for the computation of Q, K, and V.}
	\label{fig:pbr_pipeline}
\end{figure}

\textbf{Conditioning Mechanism.} 
To fully leverage multi-view information from Seed3D-MV, we design a dual-level conditioning mechanism that preserves both global appearance and local texture details from the reference image:
\begin{itemize}
\item \textbf{Global Control:} We extract global feature embeddings from the reference image using a pretrained CLIP vision encoder~\cite{radford2021learning}. These embeddings replace the original text embeddings in the diffusion model, providing high-level appearance guidance throughout generation.
\item \textbf{Local Control:} For pixel-level control, we adopt a strategy similar to ImageDream~\cite{wang2023imagedream}. Specifically, we concatenate the reference image's VAE-encoded latent with the noise latent along the channel dimension, serving as additional input to DiT blocks. To reduce computational overhead, multi-view conditioning image latents are added directly to initial noise latents and fed only into the first DiT block as initial guidance.
\end{itemize} \par

\textbf{Two-Stream Network Structure.} As established in prior work~\cite{hunyuan3d-2.1,he2025materialmvp,he2025neural}, albedo and MR exhibit significant differences in physical properties and visual characteristics. Existing methods address this through high-level architectural separation, such as separate output heads or dedicated disentanglement modules in U-Net decoders. In contrast, we propose a more fine-grained yet parameter-efficient separation mechanism.

As illustrated in Figure~\ref{fig:pbr_pipeline}, we instantiate separate projection layers for Query (Q), Key (K) and Value (V) tensors for each modality (albedo and MR) within each DiT block. After computing respective Q, K, V tensors, we concatenate latent vectors from both modalities with global image conditioning and process them through a shared full-attention module. All other DiT components, including feed-forward networks, remain shared between modalities. To distinguish modalities, we introduce learnable modality embeddings that are added to positional embeddings. Finally, two decoder heads map the processed latents to albedo and MR outputs respectively. This design effectively captures modality-specific features while significantly reducing the total number of parameters compared to using completely separate networks.

\subsubsection{Seed3D-UV}
While Seed3D-MV and Seed3D-PBR generate high-quality multi-view albedo and MR images, converting these images into complete UV texture maps presents challenges. Due to limited view coverage and self-occlusions, directly baking multi-view observations into UV space results in incomplete texture maps with missing regions. To address this, we propose Seed3D-UV, a coordinate-conditioned diffusion model for UV texture completion.

\textbf{Initial Texture Baking from Multi-view Images.}
Given the 3D mesh from the shape generation stage and multi-view material images from Seed3D-PBR, we first project each image onto the mesh surface using the corresponding camera projection matrix. For each visible surface point, we determine contributing pixels according to visibility and surface normal alignment. Following established methods~\cite{liu2024text,bensadoun2024meta}, we blend contributions from multiple views using weighted averaging based on viewing angles, assigning higher weights to views with better normal alignment.
The aggregated surface colors are then baked into a 2D UV texture map using the mesh's predefined UV parameterization~\cite{floater2005surface}. Each mesh triangle is mapped to UV space, where pixel-wise colors from overlapping views are accumulated and interpolated. However, the resulting UV map often contains incomplete regions with holes and seams, particularly in areas that are occluded across all views or only partially observed.

\textbf{Coordinate-Conditioned UV Diffusion Transformer.}
To complete the partial UV texture, we introduce a coordinate-conditioned DiT that inpaints missing regions while preserving observed content. Unlike standard image inpainting that operates in pixel space, our model leverages UV coordinate information to maintain geometric consistency with the mesh structure.
Specifically, UV coordinate maps are encoded as positional tokens and incorporated into the DiT's visual stream alongside texture tokens. This geometric conditioning guides the model to respect the UV parameterization, producing completions that align properly with mesh boundaries and existing texture content. The model learns to generate plausible texture in occluded regions by understanding both the observed pixels and their spatial relationships encoded in UV coordinates.
During inference, we condition the diffusion process on the partial UV texture obtained from multi-view baking, allowing the model to fill holes and resolve inconsistencies while maintaining coherence with visible regions. Empirically, we observe that coordinate-guided conditioning produces textures with sharper transitions at UV boundaries and better alignment with mesh geometry compared to naive inpainting approaches.

\textbf{Final Integration and Export.}
The completed UV texture from Seed3D-UV is integrated into the final asset, replacing the partial texture from multi-view baking. The resulting textured mesh, with complete albedo and metallic-roughness UV maps, is exported in standard 3D formats (\textit{e.g.}, OBJ, GLB) for downstream applications, such as rendering, animation, or scene creation.
\section{Data}

The performance of 3D generation models fundamentally depends on the scale, diversity, and quality of training data. Compared to 2D data such as images and videos, 3D data processing presents significantly greater challenges due to inherent complexity and heterogeneity. To address these challenges, we develop an automated 3D data preprocessing pipeline and scalable data infrastructure that transform vast, heterogeneous raw 3D asset collections into high-quality, diverse, and consistent datasets for training robust 3D generation models.

\subsection{Data preprocessing}
To address the inherent complexity and heterogeneity of 3D data, we design a comprehensive multi-stage preprocessing pipeline that systematically transforms raw 3D asset collections into training-ready datasets. Each stage addresses specific challenges in 3D data processing, ensuring that only high-quality assets meeting our criteria are included in the final training dataset.

\begin{figure}[htbp]
	\centering
	\includegraphics[width=\columnwidth]{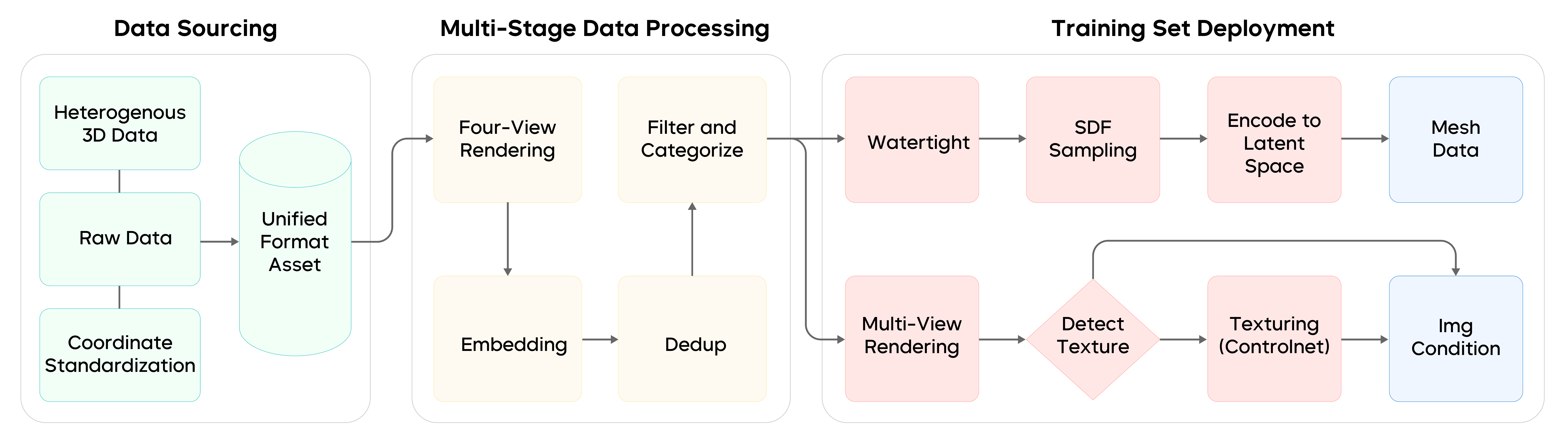}
	\caption{\textbf{Data preprocessing pipeline of Seed3D 1.0.} Our automated pipeline transforms raw 3D assets through format standardization, geometric deduplication, orientation canonization, and quality filtering, followed by multi-view rendering and mesh remeshing to produce training-ready datasets.}
	\label{fig:data_pipeline}
\end{figure}

\textbf{Diversity-Oriented Data Sourcing.} 
Our 3D data acquisition strategy prioritizes ethically and legally sourced content from diverse public repositories, licensed marketplaces, and synthetic generation platforms. We maximize coverage across critical dimensions including geometric complexity, mesh topology, object categories (\textit{e.g.}, characters, vehicles, furniture, architecture), artistic styles, material properties, and surface details. Raw collections exhibit significant heterogeneity in file formats, coordinate systems, and quality standards, often containing corrupted geometries that our pipeline addresses.

\textbf{Format Standardization and Conversion.} 
Raw 3D assets arrive in various formats such as OBJ, FBX, GLTF, PLY, and proprietary formats. We employ automated conversion tools to standardize assets into unified mesh representations, extracting geometry and material information while normalizing coordinate systems. All assets are converted to GLB format, which provides compact binary encoding and widespread compatibility across 3D applications.

\textbf{Geometric Data Deduplication.} 
3D asset collections frequently contain duplicate or near-duplicate meshes that introduce training bias and reduce dataset diversity. We develop a visual similarity-based deduplication pipeline using rendered image features and efficient nearest-neighbor search to identify and remove redundant assets. Specifically, we render each asset from four canonical viewpoints, generating RGB images and normal maps.
We employ a pretrained vision encoder~\cite{oquab2023dinov2} to extract compact representations from both modalities, concatenating features across all views to form the final mesh representation. Using FAISS~\cite{johnson2019billion} for efficient large-scale similarity search, we apply dual-threshold filtering based on cosine similarity and L2 distance to balance duplicate removal with preservation of legitimate geometric variations.

\textbf{Mesh Orientation Canonization.}
Consistent mesh orientation is crucial for effective 3D model training, as variations in object pose significantly impact model learning. We implement automated orientation canonization to standardize the spatial alignment of 3D assets.
Leveraging the same four-view renderings from the deduplication stage, we extract visual features and feed them into a trained orientation classifier that predicts canonical orientation. The predicted transformation is then applied to align the mesh to its canonical pose. This ensures that geometrically similar objects maintain consistent spatial alignment across the dataset.

\textbf{Quality Filtering with Aesthetic Scoring and VLM Assessment.} 
Raw 3D collections often contain low-quality assets with poor geometry, unrealistic proportions, or visual artifacts. We implement a two-stage quality filtering system that combines automated aesthetic evaluation with VLM-based assessment~\cite{Qwen2.5-VL}, reusing the four-view renderings from previous stages.
The first stage applies aesthetic scoring using an open-source model~\cite{aes} to evaluate visual appeal, filtering assets below a predefined threshold. The second stage employs a fine-tuned VLM for comprehensive assessment across three dimensions: (1) quality classification (unusable, usable, high-quality), (2) category identification (characters, vehicles, furniture, etc.), and (3) data type detection (synthetic \textit{vs.} real-world scanned \textit{vs.} scene-level data).
Final filtering retains only assets with acceptable aesthetic scores and usable-or-higher quality ratings, while excluding real-world scanned and scene-level data. This ensures our training dataset consists of high-quality 3D objects suitable for  foundation model training.

\textbf{Multi-View Image Rendering.} 
To bridge the gap between 3D geometry and 2D conditioning, we generate high-quality multi-view rendered images for each processed mesh using Blender's~\cite{blender2024} Cycles rendering engine. Our pipeline employs physically-based rendering with diverse lighting conditions, camera viewpoints, and material assignments to create comprehensive visual representations for model training.

For geometry generation, we render reference images from randomly sampled viewpoints with elevation angles in $[-30^\circ, 70^\circ]$ under stochastic illumination: point lights with 30\% probability or HDR environment maps with 70\% probability. For multi-view generation and PBR estimation, we sample random HDRI environments from a curated library and render normalized 3D objects from orthogonal viewpoints. Each asset is rendered to produce RGB images, normal maps, and camera coordinate maps (CCMs). For PBR training, we additionally render albedo and metallic-roughness maps, along with one fully-lit reference view to provide appearance context.
For UV texture synthesis, we unwrap 3D meshes into UV layouts using xatlas~\cite{xatlas} and bake albedo and CCMs using Blender's baking system.

\textbf{Mesh Remeshing.}
To enable valid SDF extraction for VAE training, we convert arbitrary raw meshes into watertight representations using a CUDA-based remeshing pipeline. Our approach efficiently removes internal structures while preserving external surface details through four stages: (1) voxelization using fast raster-like kernels~\cite{schwarz2010fast} with boundary marking, (2) signed distance floodfill to classify interior and exterior regions, (3) mesh extraction with threshold $\epsilon$ to preserve thin structures, and (4) final mesh generation via Dual Marching Cubes~\cite{schaefer2002dual}, with reference to the original mesh for zero-crossing normals.

\begin{figure}[htbp]
	\centering
	\includegraphics[width=\columnwidth]{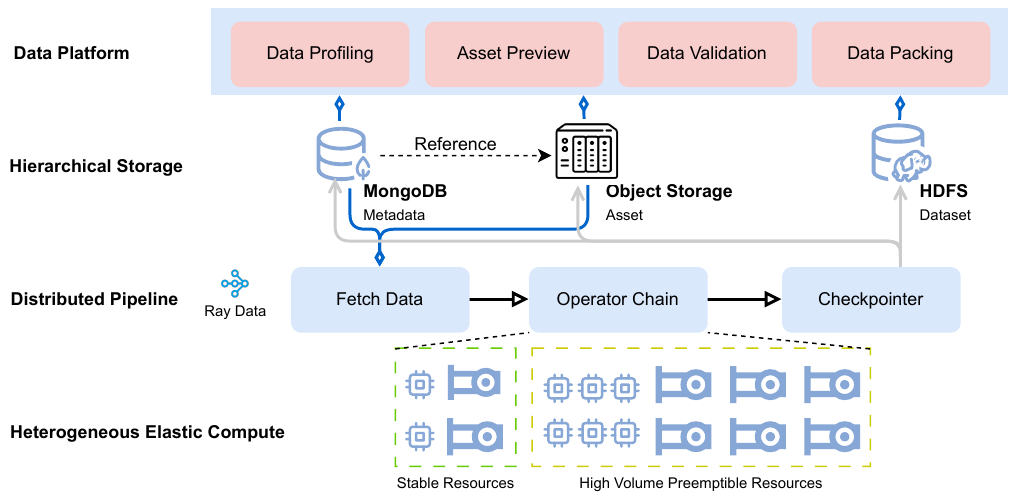}
	\caption{\textbf{Overview of Seed3D 1.0 data infrastructure.} The system integrates a web-based data platform, hierarchical storage (MongoDB, object storage, HDFS), and Ray Data distributed processing with elastic CPU/GPU resource scheduling.}
	\label{fig:data_infra}
\end{figure}

\subsection{Data Engineering Infrastructure}
To ensure scalability, traceability, and seamless integration throughout our data pipeline, we develop a comprehensive data engineering infrastructure comprising three integrated components: a centralized data management system for metadata indexing and API access, a unified storage and visualization platform for asset persistence and interactive curation, and a distributed processing infrastructure for high-throughput execution with fault tolerance.

\textbf{Data Management and Indexing.}
All metadata associated with 3D assets—including source provenance, file format, processing status, and storage paths—are indexed in a MongoDB~\cite{mongodb2024} database. Each asset is tracked throughout the pipeline via consistent metadata schemas and status flags, enabling robust querying, progress monitoring, and dataset curation.

To simplify database interactions, we implement a custom object-relational mapping (ORM) layer that exposes a standardized API for asset registration, metadata updates, and querying. This abstraction serves as the foundation for all internal automation tools and decouples preprocessing logic from backend storage systems.

\textbf{Storage and Visualization Platform.}
Raw files and intermediate outputs (\textit{e.g.}, rendered images, VLM annotations) are stored in a scalable object storage system, with asset references maintained in MongoDB and resolved at runtime via the ORM layer. This separation of metadata and content enables lightweight access and high-throughput parallel processing.

We build a web-based data platform on top of this storage infrastructure to support visual inspection and programmatic dataset operations. The platform provides filtering, tagging, thumbnail browsing, and a WebGL~\cite{webgl2011}-based 3D viewer, allowing curators and engineers to interactively explore assets, inspect rendering results, and manage asset categories.

For training data preparation, we package processed assets—including SDF samples and VAE latent codes—into training-ready bundles stored in a distributed HDFS~\cite{shvachko2010hadoop} cluster. A dedicated data packing module integrated into the web platform enables users to curate and export structured datasets based on asset categories, quality filters, or processing stages.

\textbf{Distributed Processing Infrastructure.}
We leverage Ray Data~\cite{moritz2018ray} to build a scalable distributed preprocessing pipeline that handles diverse 3D operations, including VLM-based quality assessment, multi-view rendering, and mesh remeshing. A key challenge in 3D data processing is the heterogeneous computational requirements across pipeline stages. For example, image rendering requires significant CPU resources while mesh remeshing demands GPU acceleration for intensive geometric computations. 

To address this, we deploy a custom Kubernetes~\cite{burns2016borg} operator that launches CPU and GPU pods with appropriate resource allocation for each processing stage. To maximize cost efficiency at scale, we leverage Ray Data's elasticity and fault tolerance to utilize preemptible resources from cluster idle capacity. When preemptible instances are reclaimed by higher-priority workloads, the system automatically launches replacement pods and reschedules tasks seamlessly.

Additionally, we implement strategic checkpointing after each major processing stage, enabling pipeline restart from intermediate points rather than full reprocessing. This design ensures efficient pipeline execution despite infrastructure disruptions while minimizing computational waste.

\section{Model Training}

\subsection{Geometry}

Our Seed3D-DiT training employs a three-stage progressive strategy: pre-training (PT), continued training (CT), and supervised fine-tuning (SFT). This approach enables efficient learning while progressively improving model capacity and output quality.

\textbf{Pre-Training (PT).} We train the model from scratch on low-resolution representations with 256 latent tokens to establish foundational shape generation capabilities. This stage focuses on learning fundamental geometric representations and cross-modal alignment between image conditions and 3D shapes. We use the full training dataset encompassing diverse object categories and viewing angles to ensure robust generalization.

\textbf{Continued Training (CT).} Building upon the pre-trained model, we progressively increase the latent sequence length to 4096 tokens, enabling capture of finer geometric details and surface structures. We continue training on the full dataset with enhanced data augmentation to maintain generalization performance at higher resolutions.

\textbf{Supervised Fine-Tuning (SFT).} After CT, we fine-tune the model on a curated high-quality subset with reduced learning rates to further improve generation quality, producing 3D objects with enhanced geometric accuracy and surface detail.

\subsection{Texture}
We train all texture generation models (Seed3D-MV, Seed3D-PBR, Seed3D-UV) from scratch using a two-stage approach. In the first stage, we train on the full dataset to learn comprehensive multi-view consistency and material decomposition. In the second stage, we fine-tune on a curated high-quality subset with reduced learning rates, improving output quality while maintaining robust generalization across diverse textures and materials.
\section{Training Infrastructure}
Large-scale diffusion model training requires efficient utilization of computational resources and robust failure handling mechanisms. We develop a comprehensive training infrastructure that integrates hardware-aware optimizations, memory-efficient parallelism strategies, and fault tolerance mechanisms to enable stable, high-throughput training at scale.

\subsection{Kernel Fusion}
To maximize GPU utilization, we integrate \texttt{torch.compile} with custom CUDA kernels for performance-critical operators. Through profiling analysis, we identify memory-bound operations as the primary bottleneck. We fuse multiple consecutive element-wise operations into unified kernels, reducing memory access overhead and improving arithmetic intensity. Additionally, we employ optimized libraries such as FlashAttention~\cite{dao2022flashattention} for attention computation and Apex fused optimizers for weight updates, substantially reducing computational costs. These kernel-level optimizations collectively reduce GPU idle time and improve end-to-end training throughput.

\subsection{Parallelism Strategy}
Scaling diffusion model training across multiple GPUs requires balancing communication overhead with memory efficiency. 
We employ Hybrid Sharded Data Parallelism (HSDP)~\cite{zhao2023pytorch}, which combines data parallelism within nodes and Fully Sharded Data Parallelism (FSDP) across nodes. This hierarchical approach achieves memory-efficient weight and optimizer state sharding while minimizing cross-node communication, enabling effective scaling to large cluster configurations with reduced performance degradation.

\subsection{Multi-Level Activation Checkpointing}
Memory constraints represent a fundamental bottleneck in training large diffusion transformers. While full gradient checkpointing~\cite{chen2016training} alleviates GPU memory pressure, it introduces substantial recomputation overhead during backpropagation.
To address this trade-off, we employ Multi-Level Activation Checkpointing (MLAC)~\cite{MLAC}, which balances memory usage and computational overhead. 
MLAC selectively checkpoints activations based on recomputation cost, offloading high-cost tensors to CPU memory with asynchronous prefetching to overlap memory transfers with computation. This approach achieves significant memory savings with minimal performance impact compared to full checkpointing.

\subsection{Training Stability and Fault Tolerance}
Large-scale distributed training is susceptible to hardware failures and communication disruptions. To ensure robust and reliable training execution, we implement a comprehensive stability framework combining proactive failure prevention and reactive recovery mechanisms.
Our system performs machine health checks before job launch to eliminate faulty nodes and potential stragglers. During training, we integrate flight recorder capabilities to track NCCL~\cite{nccl2017} communication patterns and identify problematic machines upon failures. 
Furthermore, we develop a centralized monitoring system that aggregates real-time performance metrics across the cluster, including Effective Training Time Ratio (ETTR), communication patterns, and GPU utilization. This provides comprehensive visibility into cluster health, enabling rapid diagnosis and resolution of bottlenecks in production training environments.
\section{Inference}
\begin{figure}[t]
	\centering
	\includegraphics[width=\columnwidth]{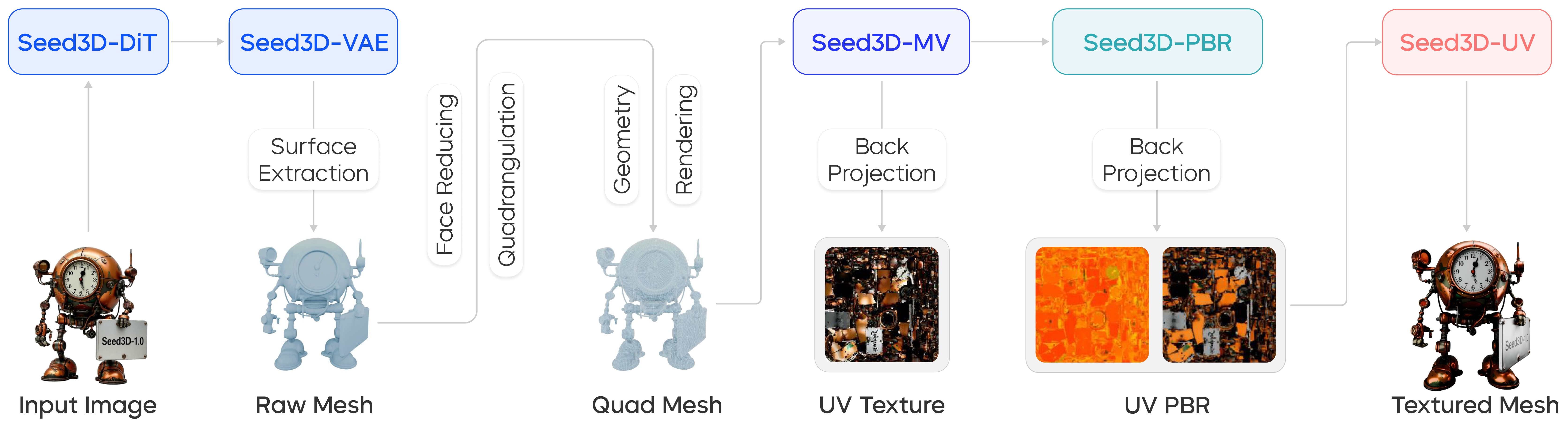}
	\caption{\textbf{Inference pipeline of Seed3D 1.0.} Given an input image, our system generates complete textured 3D assets through five sequential stages: geometry generation (Seed3D-DiT + VAE decoder), multi-view synthesis (Seed3D-MV), PBR material estimation (Seed3D-PBR), UV texture completion (Seed3D-UV), and final asset integration. The pipeline produces simulation-ready assets with watertight geometry and physically-based materials.}
	\label{fig:infer_pipeline}
\end{figure}

Figure~\ref{fig:infer_pipeline} illustrates the complete Seed3D 1.0 inference pipeline. Given an input image, our system generates a textured 3D asset through sequential multi-stage processing: geometry generation, multi-view synthesis, PBR material estimation, and UV texture completion.

\textbf{Geometry Generation.} We preprocess the input image and feed it into Seed3D-DiT to predict the 3D shape  in  latent space. The Seed3D-VAE decoder reconstructs the mesh using dual marching cubes (DMC)~\cite{schaefer2002dual}, consistent with our training pipeline.
To accelerate iso-surface extraction while preserving numerical accuracy, we employ a hierarchical extraction strategy based on quantization and spatial filtering. Specifically, we first perform coarse SDF evaluation using reduced-precision arithmetic~\cite{micikevicius2017mixed} (bfloat16) to identify candidate zero-crossing cells. Inactive cells are pruned while active cells undergo full-precision (float32) evaluation. This substantially reduces computation while maintaining mesh fidelity. For gradient estimation required by DMC vertex placement, we leverage analytical gradients from the VAE's SDF decoder via auto-differentiation~\cite{baydin2017automatic}. The extracted mesh then undergoes retopology and UV unwrapping~\cite{floater2005surface} for subsequent material generation.

\textbf{Multi-View Generation and Initial Texturing.} Using the generated mesh and input image, Seed3D-MV produces multi-view consistent RGB images. These images are back-projected onto the mesh surface and baked into UV space, producing partial UV textures. Due to limited viewpoints and occlusions, the resulting UV maps contain incomplete regions that require subsequent enhancement.

\textbf{Material Estimation.} Seed3D-PBR decomposes the multi-view images into albedo and metallic-roughness components. These PBR maps are baked into UV space using the same projection method, providing physically-based material properties for realistic rendering.

\textbf{Texture Completion.} To complete the partial UV textures, we feed the incomplete albedo and MR UV maps into Seed3D-UV for inpainting. This diffusion-based model generates spatially coherent textures using coordinate conditioning to maintain geometric consistency.

\textbf{Final Asset Integration.} The completed texture maps—albedo, metallic, and roughness—are integrated with the mesh to produce the final 3D asset. The resulting asset features watertight, manifold geometry with optimized topology, suitable for rendering, simulation, and interactive applications. Assets are exported in standard formats (OBJ, GLB) for broad compatibility.

\section{Model Performance}

We conduct comprehensive evaluations comparing Seed3D 1.0 with  state-of-the-art methods on both geometry and texture generation tasks. Our evaluation includes quantitative benchmarks, qualitative analysis, and user studies to assess generation quality across different aspects.

\subsection{Comparisons}
\subsubsection{Geometry Generation}

\begin{table}[h]
\centering
\begin{tabular}{l|c|c|c|c}
\hline
Models & ULIP-T ($\uparrow$) & ULIP-I ($\uparrow$) & Uni3D-T ($\uparrow$) & Uni3D-I ($\uparrow$) \\
\hline
TRELLIS~\cite{trellis} & $0.0951 \pm 0.0608$ & $0.1686 \pm 0.0826$ & $0.2786 \pm 0.0671$ & $0.3754 \pm 0.0713$ \\
TripoSG~\cite{li2025triposg} & $0.1312 \pm 0.0574$ & $0.2460 \pm 0.0554$ & $0.2657 \pm 0.0652$ & $0.3870 \pm 0.0671$ \\
Step1X-3D~\cite{li2025step1x} & $0.1316 \pm 0.0573$ & $0.2441 \pm 0.0527$ & $0.2709 \pm 0.0625$ & $0.3837 \pm 0.0687$ \\
Direct3D-S2~\cite{wu2025direct3ds2gigascale3dgeneration} & $0.1203 \pm 0.0555$ & $0.2191 \pm 0.0572$ & $0.2571 \pm 0.0582$ & $0.3497 \pm 0.0697$ \\
Hunyuan3D-2.1~\cite{hunyuan3d-2.1} & $0.1283 \pm 0.0580$ & $0.2376 \pm 0.0593$ & $0.2575 \pm 0.0672$ & $0.3709 \pm 0.0769$ \\
\textbf{Seed3D 1.0} & $\textbf{0.1319} \pm \textbf{0.0572}$ & $\textbf{0.2536} \pm \textbf{0.0432}$ & $\textbf{0.2800} \pm \textbf{0.0634}$ & $\textbf{0.3999} \pm \textbf{0.0610}$ \\
\hline
\end{tabular}
\caption{\textbf{Quantitative comparison for geometry generation.} Seed3D 1.0 achieves state-of-the-art performance across all metrics.}
\label{tab:shape_generation}
\end{table}

\begin{figure}[htbp]
	\centering
	\includegraphics[width=0.88\columnwidth]{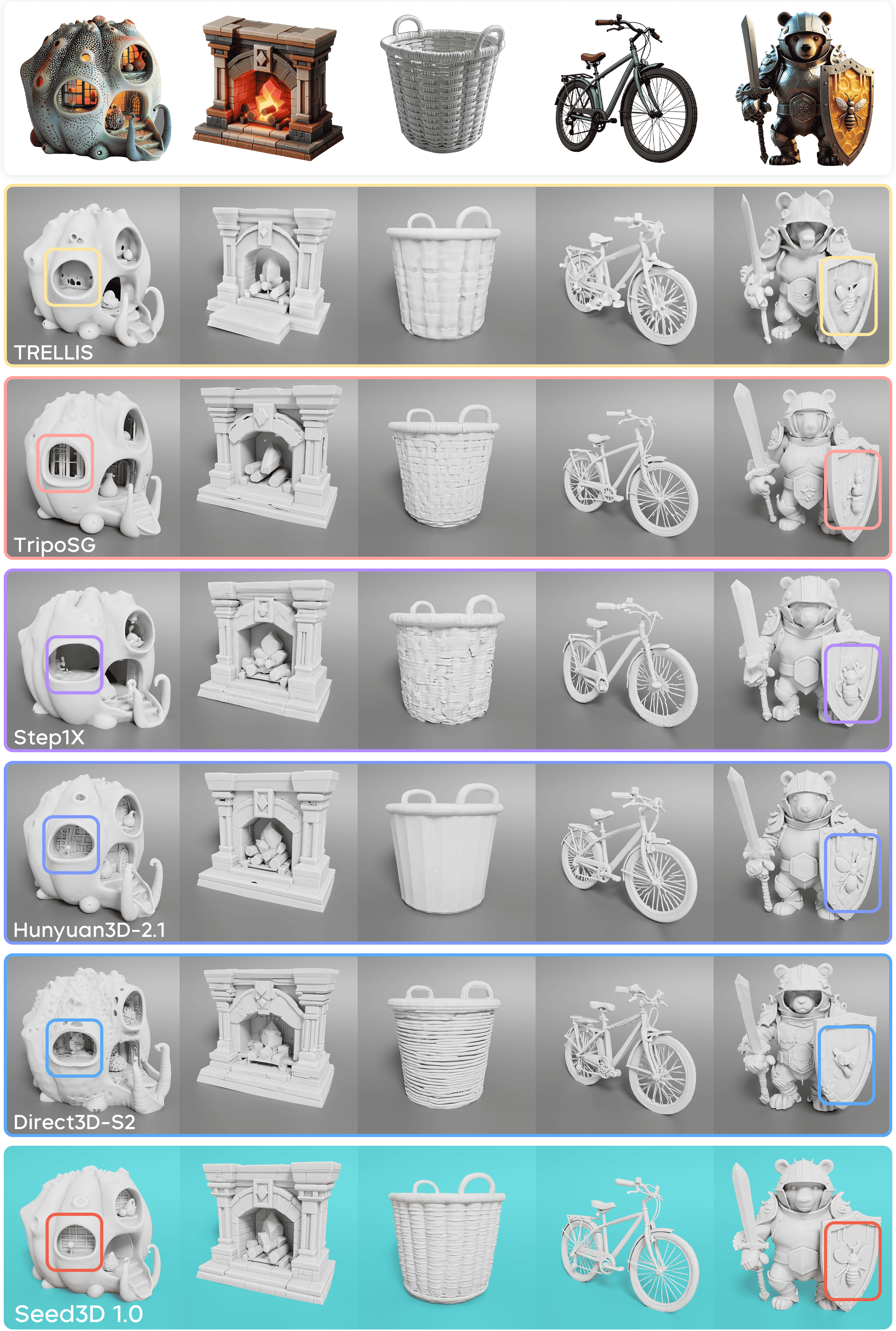}
	\caption{\textbf{Qualitative comparisons of geometry generation.} Seed3D 1.0 produces meshes with finer geometric details and better structural accuracy compared to baseline methods. Best viewed at 8$\times$ zoom.}
	\label{fig:geoperform}
\end{figure}

\textbf{Experimental Setup.} We evaluate our 1.5B-parameter Seed3D-DiT on single-image to 3D mesh task. We compare against state-of-the-art open-source methods: TRELLIS~\cite{trellis}, TripoSG~\cite{li2025triposg}, Step1X-3D~\cite{li2025step1x}, Direct3D-S2~\cite{wu2025direct3ds2gigascale3dgeneration}, and Hunyuan3D-2.1~\cite{hunyuan3d-2.1}.

\textbf{Evaluation Protocol.} We evaluate on a test set of 1,000 images covering diverse object categories (characters, furniture, animals, etc) and artistic styles (realistic, cartoon, gaming, etc). We employ ULIP~\cite{xue2024ulip} and Uni3D~\cite{zhou2023uni3d} models to measure similarity between generated meshes and input images. For each mesh, we sample 8,192 surface points and compute ULIP-I/ULIP-T and Uni3D-I/Uni3D-T scores using VLM-generated captions~\cite{Qwen2.5-VL} as text conditioning.

\textbf{Quantitative Results.} Table~\ref{tab:shape_generation} shows Seed3D 1.0 achieves the highest scores across all metrics. Notably, our 1.5B model outperforms the 3B Hunyuan3D-2.1, demonstrating the effectiveness of our model architecture and training approach. The strong ULIP-I and Uni3D-I scores indicate excellent alignment between generated geometry and input images.

\textbf{Qualitative Analysis.} The geometry generation performance of our Seed3D 1.0 can be further verified by the qualitative results. As shown in Figure~\ref{fig:geoperform}, our method generates superior results compared to baseline methods in terms of geometric detail preservation, structural accuracy, and overall shape fidelity. Visual inspection confirms that Seed3D captures intricate features such as the complex structures of architectural elements, the fine textures of woven baskets, and the precise geometry of mechanical objects like bicycles.

\subsubsection{Texture Generation}

\textbf{Experimental Setup.} We evaluate our multi-view generation and PBR estimation models using both image and geometry conditioning. We compare against open-source methods: MVPainter~\cite{shao2025mvpainteraccuratedetailed3d}, Hunyuan3D-Paint~\cite{hunyuan3d-2.0}, UniTEX~\cite{liang2025UnitTEX}, MV-Adapter~\cite{huang2024mvadapter}, Pandora3d~\cite{yang2025pandora3d}, and Hunyuan3D 2.1~\cite{hunyuan3d-2.1}.

\textbf{Evaluation Protocol.} We employ established metrics including CLIP~\cite{radford2021learning}-based Fréchet Inception Distance (CLIP-FID)~\cite{heusel2017gans}, Learned Perceptual Image Patch Similarity (LPIPS)~\cite{zhang2018perceptual}, CLIP Maximum-Mean Discrepancy (CMMD), and CLIP-Image Similarity (CLIP-I).

\begin{table}[h]
\centering
\begin{tabular}{l|c|c|c|c}
\hline
Method & CLIP-FID ($\downarrow$) & CMMD ($\downarrow$) & CLIP-I ($\uparrow$) & LPIPS ($\downarrow$) \\
\hline
MVPainter~\cite{shao2025mvpainteraccuratedetailed3d} & 31.7290 & 0.3254 & 0.8903  & 0.1420 \\
Hunyuan3D-Paint~\cite{hunyuan3d-2.0} & 18.8625 & 0.0825 & 0.9206  & 0.1162 \\
UniTEX~\cite{liang2025UnitTEX}       & 18.3285 & 0.0873 & 0.9230  & 0.1078 \\
MV-Adapter~\cite{huang2024mvadapter} & 11.6920 & 0.0312 & 0.9399  & 0.1012 \\
\textbf{Seed3D 1.0}  & \textbf{9.9752} & \textbf{0.0231} & \textbf{0.9484} & \textbf{0.0891} \\
\hline
\end{tabular}
\caption{\textbf{Quantitative comparison for multi-view generation.} Seed3D 1.0 achieves state-of-the-art performance across all metrics.}
\label{tab:mv}
\end{table}

\begin{table}[h]
\centering
\begin{tabular}{l|c|c|c|c}
\hline
Method & CLIP-FID ($\downarrow$) & CMMD ($\downarrow$) & CLIP-I ($\uparrow$) & LPIPS ($\downarrow$) \\
\hline
Pandora3d~\cite{yang2025pandora3d} & 37.7028 & 0.3650 & 0.8868  & 0.1229 \\
MVPainter~\cite{shao2025mvpainteraccuratedetailed3d} & 40.6763 & 0.4145 & 0.8724  & 0.1274 \\
Hunyuan3D-2.1~\cite{hunyuan3d-2.1}       & 36.3484 & 0.3026 & 0.8828  & 0.1318 \\
\textbf{Seed3D 1.0}        & \textbf{31.5984} & \textbf{0.2795} & \textbf{0.9000}  & \textbf{0.1153} \\
\hline
\textcolor{gray}{\textbf{Seed3D 1.0\(^*\)}}  & \textcolor{gray}{\textbf{23.3919}} & \textcolor{gray}{\textbf{0.2191}} & \textcolor{gray}{\textbf{0.9310}} & \textcolor{gray}{\textbf{0.0843}} \\
\hline
\end{tabular}
\caption{\textbf{Quantitative comparison for PBR material generation.} Seed3D 1.0 achieves the best performance. \textcolor{gray}{\textbf{Seed3D 1.0\(^*\)}} uses ground-truth multi-view images, demonstrating the upper-bound performance when decoupled from multi-view generation errors.}
\label{tab:pbr}
\end{table}

\begin{figure}[htbp]
\centering
    \includegraphics[width=0.95\columnwidth]{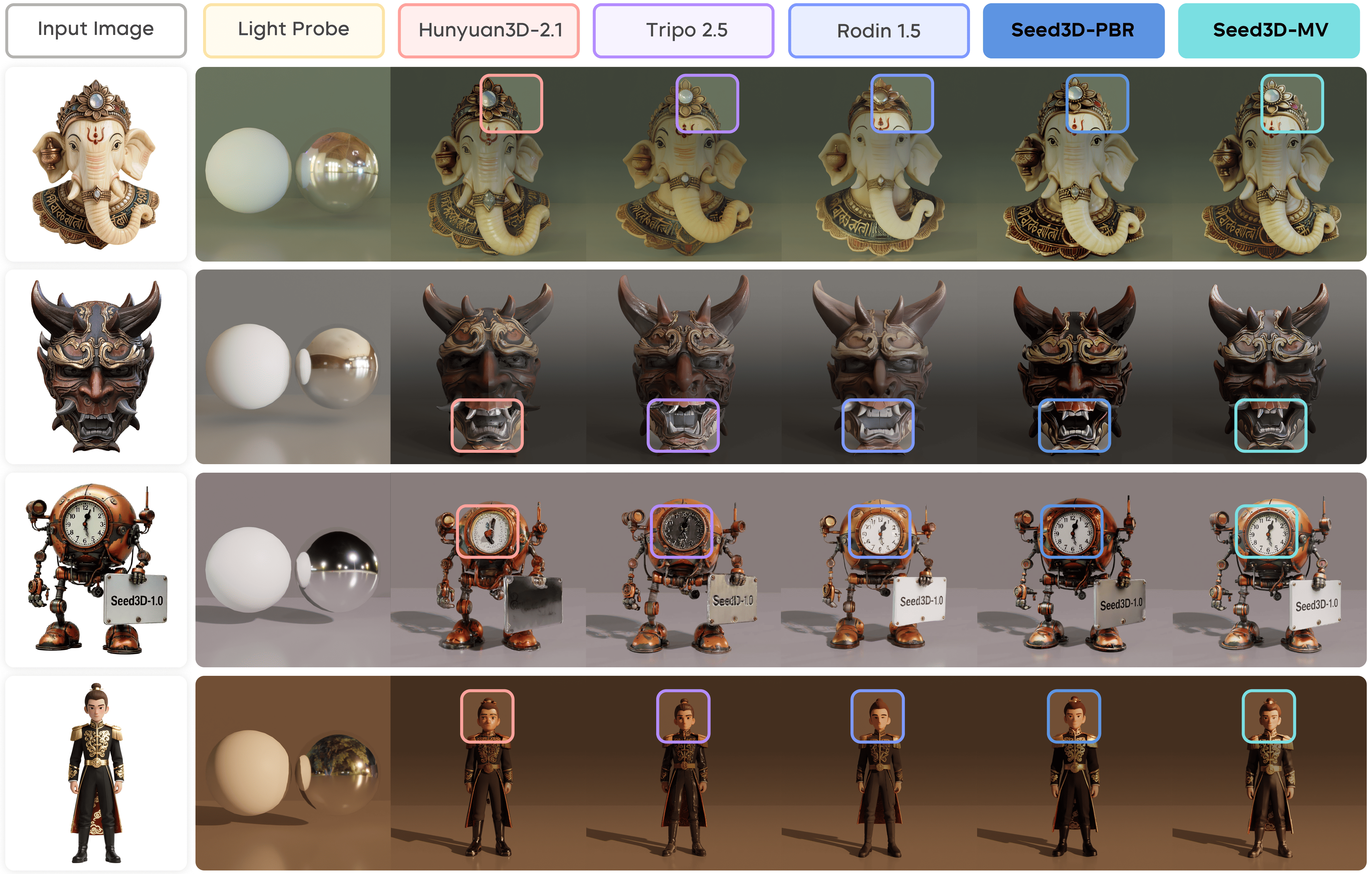}
	\caption{\textbf{Qualitative comparison of texture generation.} Red boxes highlight improvements in fine-grained detail preservation, text clarity, and material quality. Best viewed at 8$\times$ zoom.}
	\label{fig:qual_tex}
\end{figure}

\begin{figure}[htbp]
    \centering
    \scalebox{0.95}{
    \begin{minipage}{\textwidth}
        \begin{subfigure}[b]{0.64\textwidth}
            \centering
            \includegraphics[width=\textwidth]{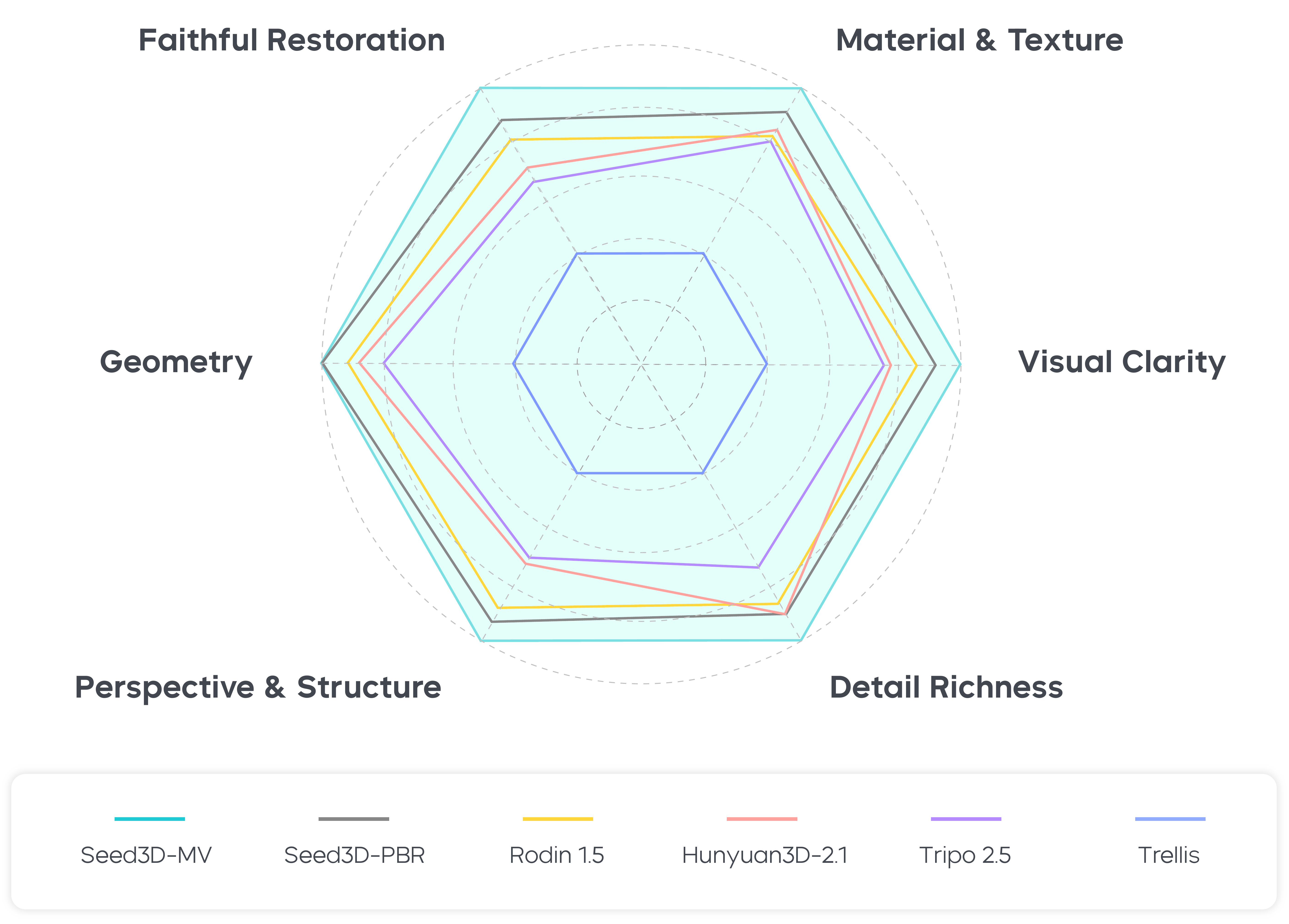}
            \caption{\textbf{User study} comparing Seed3D against baseline methods across multiple quality dimensions.}
            \label{fig:user_study}
        \end{subfigure}
        \hfill
        \begin{subfigure}[b]{0.34\textwidth}
            \centering
            \includegraphics[width=\textwidth]{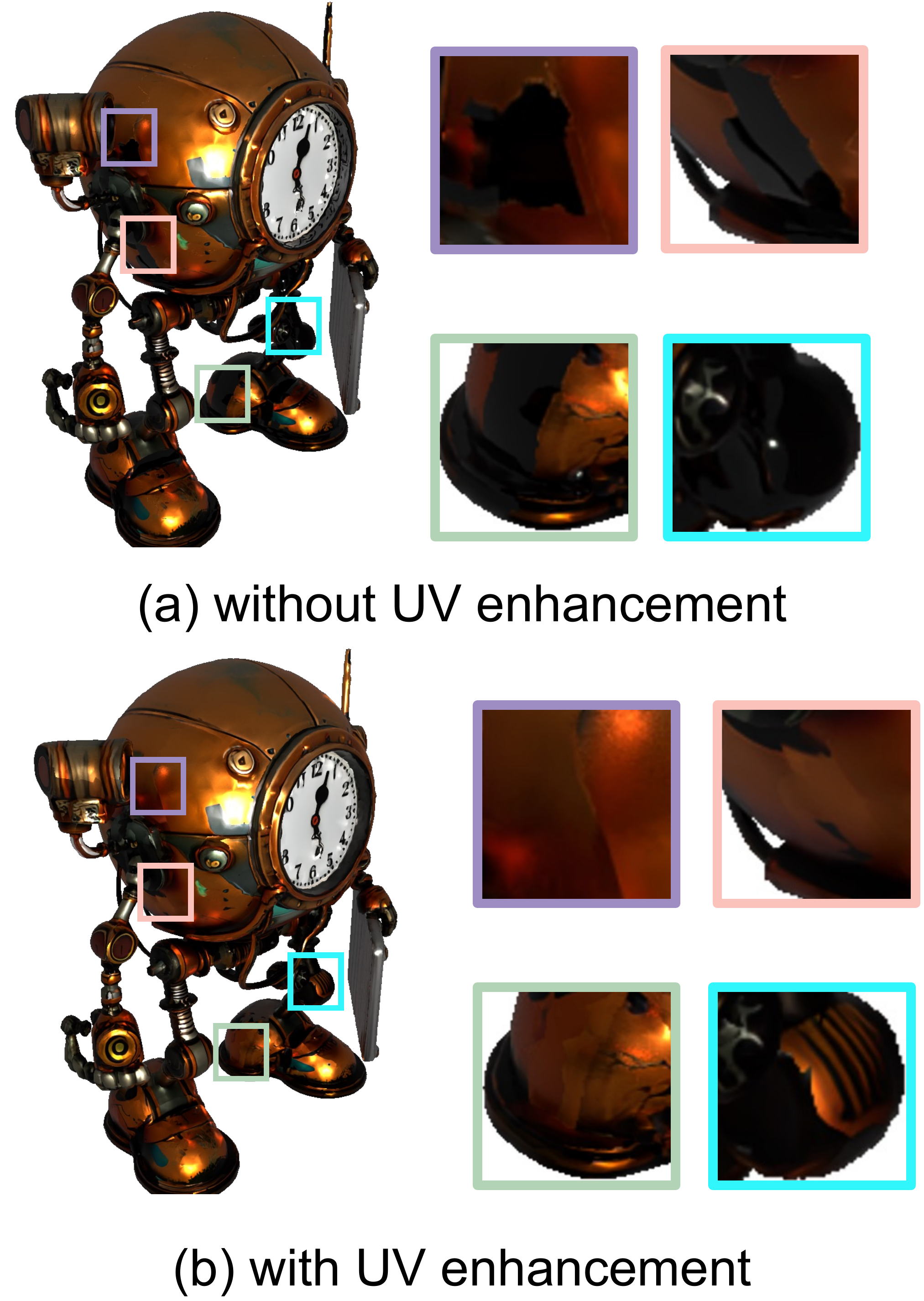}
            \caption{\textbf{Ablation of UV enhancement.} Seed3D-UV inpaints missing textures caused by self-occlusion.}
            \label{fig:qual_uv}
        \end{subfigure}
    \end{minipage}
    }
    \caption{\textbf{User study and ablation analysis.} Our method demonstrates superior performance in human evaluation and benefits significantly from UV texture completion.}
    \label{fig:main}
\end{figure}

\textbf{Quantitative Results.} Table~\ref{tab:mv} shows Seed3D-MV achieves state-of-the-art performance across all multi-view generation metrics. Table~\ref{tab:pbr} presents PBR estimation results, where we use multi-view images generated by Seed3D-MV as input for fair comparison. Seed3D-PBR demonstrates the best performance among all methods. We also report results using ground-truth multi-view images (Seed3D 1.0\(^*\)), which represent the upper-bound performance when decoupled from multi-view generation errors, showing significant improvements with higher-quality inputs.

\textbf{Qualitative Analysis.} Figure~\ref{fig:qual_tex} provides qualitative comparisons demonstrating Seed3D 1.0's superior texture and material quality. Our method shows notable improvements in preserving fine-grained details from reference images and rendering clear text elements.
Seed3D 1.0 maintains strong alignment with reference images, particularly for detailed visual features. As shown in the last row of Figure~\ref{fig:qual_tex}, baseline methods tend to lose reference fidelity, while Seed3D 1.0 accurately generates fine details such as facial features and textile patterns. The generated PBR materials exhibit realistic surface properties, including appropriate metallic reflectance and skin subsurface scattering, contributing to photorealistic rendering results.

The superiority of our approach is also evident across other challenging scenarios. In the steampunk clock example (the third row in Figure~\ref{fig:qual_tex}), while other methods produce blurred details, Seed3D 1.0 maintains sharp clarity for fine textual elements like numbers on clock face and mechanical components. This demonstrates exceptional preservation of high-frequency texture details crucial for realistic 3D generation.

\textbf{UV Enhancement Analysis.} 
Figure~\ref{fig:qual_uv} demonstrates the effectiveness of Seed3D-UV. Without UV enhancement, back-projection from limited viewpoints results in incomplete texture maps with missing regions due to self-occlusion. Seed3D-UV successfully inpaints these incomplete regions, producing complete and spatially coherent UV textures.

\subsection{User Study} 
We conduct a user study with 14 human evaluators to assess generation quality across 43 diverse test images. Evaluators compare 6 methods across multiple dimensions: visual clarity, faithful restoration, geometry quality, perspective \& structure accuracy, material \& texture realism, and detail richness. As shown in Figure~\ref{fig:user_study}, Seed3D 1.0 receives consistently higher ratings across all dimensions, with particularly strong performance in geometry and material quality.
\begin{figure}[ht]
    \centering
    \includegraphics[width=\linewidth]{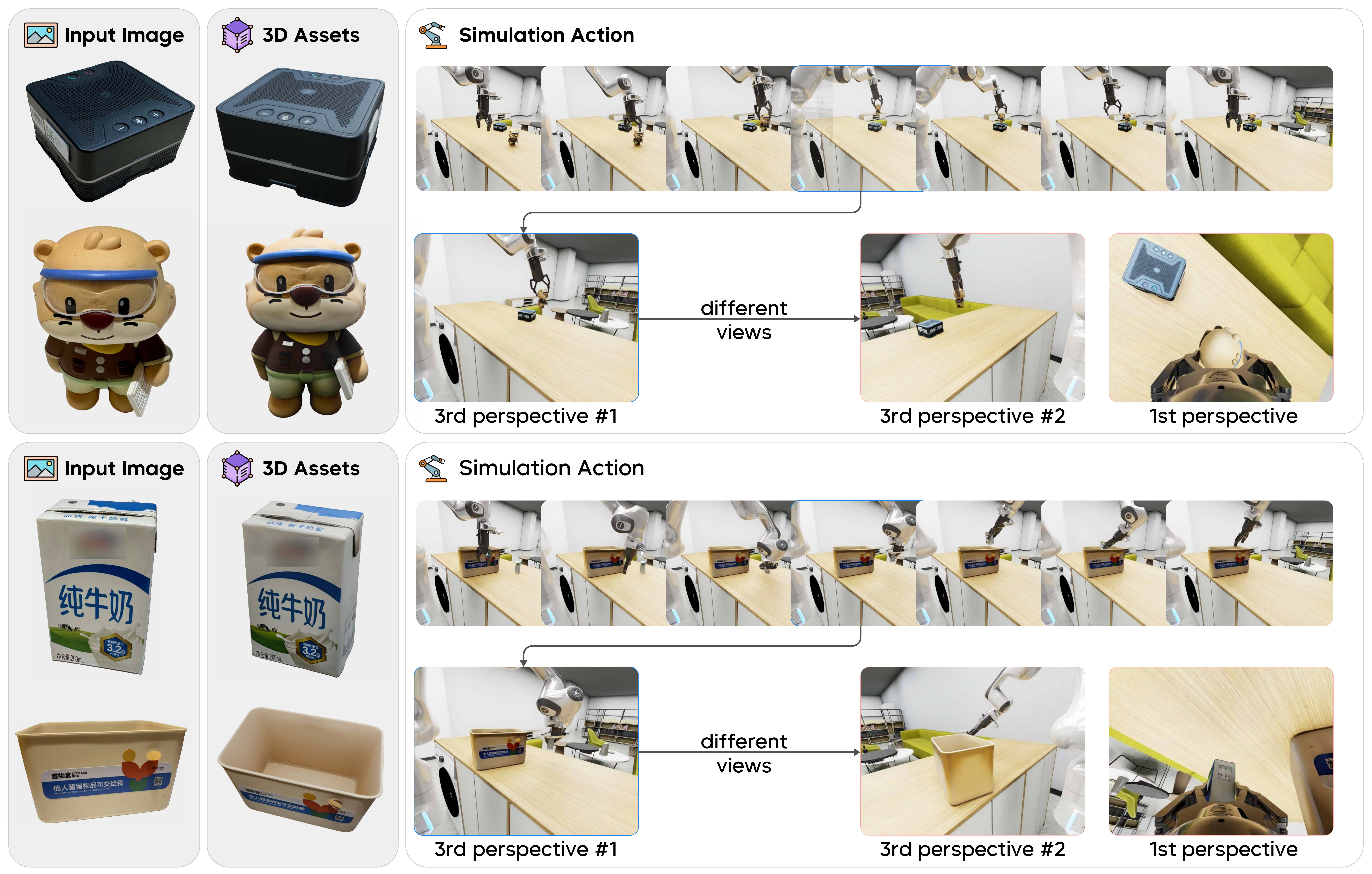}
    \caption{\textbf{Simulation-ready asset generation for robotics.} Seed3D 1.0 generates physics-compatible 3D assets from single images, including electronic devices, toys, storage containers, and household items. Generated assets can be easily integrated into Isaac Sim for robotic manipulation tasks, maintaining geometric accuracy and material fidelity across multiple viewpoints for realistic grasping and manipulation simulations. Best viewed with 8$\times$ zoom.}
    \label{fig:app1}
\end{figure}

\section{Application}

\subsection{Simulation-ready Generation}

Figure~\ref{fig:app1} demonstrates Seed3D 1.0's capability to generate assets suitable for physics-based simulation. Given a single input image, our system produces 3D assets that can be integrated into NVIDIA Isaac Sim~\cite{makoviychuk2021isaac} for robotic manipulation testing. To import the assets into the simulator, we utilize VLM~\cite{Qwen2.5-VL} to estimate the scale of each asset and adjust them to real-world dimensions.
Isaac Sim automatically generates collision meshes from the watertight, manifold geometry and applies default material properties (\textit{e.g.}, friction), enabling immediate physics simulation without manual tuning.

We conduct robotic manipulation experiments including grasping and multi-object interactions within Isaac Sim. The physics engine provides real-time feedback on contact forces, object dynamics, and manipulation outcomes. Assets generated by Seed3D 1.0 preserve fine geometric details essential for realistic contact simulation—for example, toys and electronic devices maintain accurate surface features crucial for grasp planning. Combined with comprehensive physics simulation, these environments offer three key benefits for embodied AI development: (1) scalable generation of training data through diverse manipulation scenarios, (2) interactive learning via physics feedback on action consequences, and (3) diverse multi-view, multi-modal observation data that enables comprehensive evaluation benchmarks for vision-language-action (VLA) models.

\subsection{Scene Generation}
\begin{figure}[h]
    \centering
    \includegraphics[width=\linewidth]{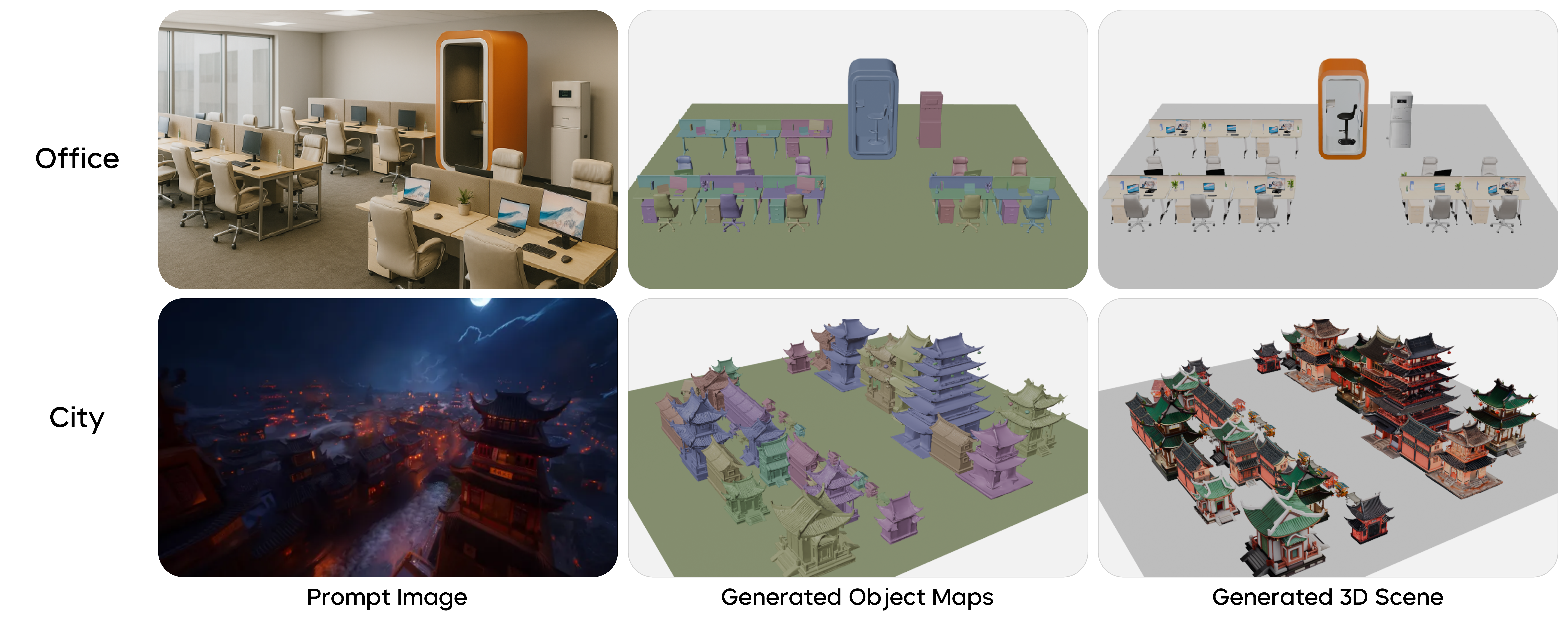}
    \caption{\textbf{Factorized scene generation.} Given prompt images (left), our system employs VLMs to generate object layout maps specifying positions, scales, and orientations (center). Individual objects are then generated and assembled into complete 3D scenes (right). Examples demonstrate coherent scene generation for office and traditional urban environments.}
    \label{fig:appscene}
\end{figure}

Seed3D 1.0 extends to scene-level generation through a factorized approach. As demonstrated in Figure~\ref{fig:appscene}, given input prompt images, we employ a VLM to identify objects and infer their spatial relationships, generating layout maps that specify object scales, positions, and orientations. The system then generates geometry and texture for each object individually. The final scene is assembled by positioning objects according to the predicted layout, enabling coherent scene generation across diverse environments from indoor offices to urban architectural scenes.
\section{Conclusion}
We present Seed3D 1.0, a foundation model for generating simulation-ready 3D assets from single images. Our system generates high-quality assets with detailed geometry, photorealistic textures, and physically-based materials through four integrated components: Seed3D-DiT for geometry generation, Seed3D-MV for multi-view synthesis, Seed3D-PBR for material decomposition, and Seed3D-UV for texture completion, supported by scalable data infrastructure and optimized training systems.
Experimental results demonstrate state-of-the-art performance across geometry and texture generation benchmarks. Quantitative evaluations show our 1.5B parameter geometry generation model achieves superior results compared to larger baseline methods, while comprehensive user studies validate generation quality across visual clarity, geometric accuracy, and material realism.
A key strength of Seed3D 1.0 is generating physics-compatible assets that integrate directly into simulation environments. Generated meshes maintain watertight, manifold geometry, enabling immediate deployment in physics engines such as Isaac Sim without manual preprocessing. We demonstrate practical applications in robotic manipulation simulation, where these assets support scalable training data generation and comprehensive evaluation benchmarks for VLA models. Our approach also extends to scene-level generation through factorized composition, assembling individual objects into coherent environments.
By enabling scalable generation of simulation-ready 3D content, Seed3D 1.0 advances the development of physics-based world simulators for embodied AI, providing a foundation for training embodied agents capable of realistic physical interaction.

\clearpage

\bibliographystyle{plainnat}
\bibliography{main}

\clearpage

\beginappendix
\section{Contributions and Acknowledgments}
\label{sec:contributions}

All contributors of Seed3D are listed in alphabetical order by their last names.

\subsection{Core Contributors}
Jiashi Feng, Xiu Li, Jing Lin, Jiahang Liu, Gaohong Liu, Weiqiang Lou, Su Ma, Guang Shi, Qinlong Wang, Jun Wang, Zhongcong Xu, Xuanyu Yi, Zihao Yu, Jianfeng Zhang, Yifan Zhu

\subsection{Contributors}
Rui Chen, Jinxin Chi, Zixian Du, Li Han, Lixin Huang, Kaihua Jiang, Yuhan Li, Guan Luo, Shuguang Wang, Qianyi Wu, Fan Yang, Junyang Zhang, Xuanmeng Zhang

\subsection{Acknowledgments}
Hengkai Guo, Xiaoyang Guo, Liang Han, Xu Han, Junrui Hao, Minghan Qin, Huiyao Shu, Wanxing Wang, Zhuolin Zheng, Eddie Zhou, Jiaqing Zhou

\end{document}